\documentclass[12pt]{article}

\usepackage{graphics}
\usepackage{graphicx}
\usepackage{amssymb}
\usepackage{amsmath}
\usepackage{amsfonts}
\usepackage{cite}

\newif\ifFIG
\FIGtrue

\begin{document}

\begin{titlepage}
\begin{center}

{\Large
 \bf{Inverted-sandwich-type and open-lantern-type dinuclear transition metal complexes: 
 theoretical study of chemical bonds by electronic stress tensor}
 }

\vskip .45in

{\large
Kazuhide Ichikawa$^1$, Ayumu Wagatsuma$^1$, Yusaku I. Kurokawa$^2$, \\
Shigeyoshi Sakaki$^3$ and Akitomo Tachibana$^{*1}$}

\vskip .45in

{\em
$^1$Department of Micro Engineering, Kyoto University, 
Kyoto, 606-8501, Japan\\
$^2$Quantum Chemistry Research Institute,
Kyoto, 615-8245 Japan \\
$^3$Fukui Institute for Fundamental Chemistry Organization, Kyoto University,
Kyoto, 606-8103 Japan   \\
}

\vskip .45in
{\tt E-mail: akitomo@scl.kyoto-u.ac.jp}

\end{center}

\vskip .4in

\date{\today}

\begin{abstract}
We study the electronic structure of two types of transition metal complexes, the inverted-sandwich-type and open-lantern-type, by the electronic stress tensor. 
In particular, the bond order $b_\varepsilon$ measured by the energy density which is defined from the electronic stress tensor is studied and compared with the conventional MO based bond order. 
We also examine the patterns found in the largest eigenvalue of the stress tensor and corresponding eigenvector field, the ``spindle structure" and ``pseudo-spindle structure". 
As for the inverted-sandwich-type complex, our bond order $b_\varepsilon$ calculation shows that relative strength of the metal-benzene bond among V, Cr and Mn complexes is ${\rm V > Cr > Mn}$ which is consistent with the MO based bond order.
As for the open-lantern-type complex, we find that our energy density based bond order can properly describe the relative strength of Cr--Cr and Mo--Mo bonds by the surface integration of the energy density over the ``Lagrange surface" which can take into account the spatial extent of the orbitals. 
\end{abstract}

{Wave function analysis; Theory of chemical bond; Stress tensor; Transition metal complexes}

\end{titlepage}

\setcounter{page}{1}

\section{Introduction} \label{sec:intro}
The stress tensors in quantum systems have been investigated for many years, including one of the earliest quantum mechanics papers \cite{Schrodinger1927, Pauli, Epstein1975, Bader1980, Bamzai1981a, Nielsen1983, Nielsen1985, Folland1986a, Folland1986b, Godfrey1988, Filippetti2000, Tachibana2001, Pendas2002, Rogers2002, Tachibana2004, Tachibana2005, Morante2006, Tao2008, Ayers2009, Tachibana2010}. 
The stress tensors in general are widely used for description of internal forces of matter in various fields of science such as mechanical engineering and material science. 
As for the stress tensors in quantum mechanics context, we can find several different definitions and applications in the literature.
For example, Ref.~\cite{Nielsen1983} and followers focus on the stress tensor which is associated with forces on nuclei. 
In contrast, the one we consider in this paper is the electronic stress tensor, which is associated with effects caused by internal forces acting on electrons in molecules, following Ref.~\cite{Tachibana2001}. 
This electronic stress tensor has been used to investigate chemical bonds and reactions and many interesting properties have been discovered \cite{Tachibana2001,Tachibana2004,Tachibana2005, Szarek2007, Szarek2008, Szarek2009, Ichikawa2009a, Ichikawa2009b,Tachibana2010, Ichikawa2010a, Ichikawa2010b}. 

Among them, it is shown that the energy density can be defined from the electronic stress tensor. 
Using this energy density, new definition of bond order is proposed \cite{Szarek2007}.
So far, this stress-tensor-based bond order is applied to $s$-block and $p$-block compounds in Refs.~\cite{Szarek2007,Szarek2008,Ichikawa2010b} and found to have reasonable features. 
Then, next question is whether this bond order would work well for $d$-block compounds.

In this paper, we wish to address this issue using two types of transition metal complexes.
The first one is the inverted-sandwich-type dinuclear transition metal complexes and the second one is the open-lantern-type dinuclear transition metal complexes.
Based on the electronic structures that are thoroughly investigated in Refs.~\cite{Kurokawa2010} and \cite{Kurokawa2009}, we study the electronic stress tensor of these molecules. Our special attention is given to chemical bonds between metal atoms and benzene for the former and those between transition metals for the latter. 

This paper is organized as follows. In the next section, we briefly explain the electronic stress tensor and related values including the definition of our bond order. 
In Sec.~\ref{sec:results}, we discuss our results of the electronic stress tensor analysis.
Section~\ref{sec:istc} is for the inverted-sandwich-type dinuclear transition metal complexes and Sec.~\ref{sec:oltc} is for the open-lantern-type dinuclear transition metal complexes.
We summarize our paper in Sec.~\ref{sec:summary}.

\section{Theory and calculation methods} \label{sec:calc}

In the following section, we use quantities derived from the electronic stress tensor to analyze chemical bonds of transition metal complexes. This method based on Regional Density Functional Theory and Rigged Quantum Electrodynamics \cite{Tachibana2001,Tachibana2004,Tachibana2005,Tachibana2010} provides useful quantities to investigate chemical bonding such as new definition of bond order \cite{Szarek2007,Szarek2008,Szarek2009}. We briefly describe them below.
(For other studies of quantum systems with the stress tensor in a somewhat different context, see Refs.~\cite{Bader1980,Nielsen1983,Nielsen1985,Folland1986a,Folland1986b,Godfrey1988,Filippetti2000,Pendas2002,Rogers2002,Morante2006,Tao2008,Ayers2009}.
See also Refs.~\cite{Ayers2002,Anderson2010} for related discussion on energy density.)

The basic quantity in this analysis is the electronic stress tensor density $\overleftrightarrow{\tau}^{S}(\vec{r})$ whose components are given by
\begin{eqnarray} 
\tau^{Skl}(\vec{r}) &=& \frac{\hbar^2}{4m}\sum_i \nu_i
\Bigg[\psi^*_i(\vec{r})\frac{\partial^2\psi_i(\vec{r})}{\partial x^k \partial x^l}-\frac{\partial\psi^*_i(\vec{r})}{\partial x^k} \frac{\partial\psi_i(\vec{r})}{\partial x^l} \nonumber\\
& & \hspace{4cm} +\frac{\partial^2 \psi^*_i(\vec{r})}{\partial x^k \partial x^l}\psi_i(\vec{r}) -\frac{\partial \psi^*_i(\vec{r})}{\partial x^l}\frac{\partial \psi_i(\vec{r})}{\partial x^k}\Bigg],
\label{eq:stress}
\end{eqnarray}
where $\{k, l\} = \{1, 2, 3\}$, $m$ is the electron mass, $\psi_i(\vec{r})$ is the $i$th natural orbital and $\nu_i$ is its occupation number.

By taking a trace of $\overleftrightarrow{\tau}^{S}(\vec{r})$, we can define energy density of the quantum system at each point in space. The energy density $\varepsilon_\tau^S(\vec{r})$ is given by 
\begin{eqnarray}
\varepsilon_\tau^S(\vec{r}) = \frac{1}{2} \sum_{k=1}^3 \tau^{Skk}(\vec{r}).
\end{eqnarray}
We note that, by using the virial theorem, integration of $\varepsilon_\tau^S(\vec{r})$ over whole space gives usual total energy $E$ of the system: $\int \varepsilon_\tau^S(\vec{r}) d\vec{r} = E$.


Now, we define bond orders as this energy density $\varepsilon_\tau^S(\vec{r})$ 
 at ``Lagrange point" between the two atoms \cite{Szarek2007}. The Lagrange point $\vec{r}_L$ is the point where the tension density $\vec{\tau}^S(\vec{r})$ given by the divergence of the stress tensor 
\begin{eqnarray} 
\tau^{S k}(\vec{r})&=&  \sum_l \partial_l  \tau^{Skl}(\vec{r}) \nonumber \\
&=&\frac{\hbar^2}{4m}\sum_i \nu_i
\Bigg[\psi^*_i(\vec{r})\frac{\partial \Delta\psi_i(\vec{r})}{\partial x^k}-\frac{\partial\psi^*_i(\vec{r})}{\partial x^k} \Delta\psi_i(\vec{r}) \nonumber\\
& & \hspace{4cm} +\frac{\partial \Delta\psi^*_i(\vec{r})}{\partial x^k}\psi_i(\vec{r}) -\Delta \psi^*_i(\vec{r}) \frac{\partial \psi_i(\vec{r})}{\partial x^k}\Bigg],
\label{eq:tension}
\end{eqnarray}
vanishes. Namely, $\tau^{S k}(\vec{r}_L)=0$. $\vec{\tau}^S(\vec{r})$ is the expectation value of the tension density operator $\Hat{\vec{\tau}}^S(\vec{r})$, which cancels the Lorentz force density operator $\Hat{\vec{L}}(\vec{r})$ in the equation of motion for stationary state \cite{Tachibana2001}. Therefore, we see that $\vec{\tau}^S(\vec{r})$ expresses purely quantum mechanical effect and it has been proposed that this stationary point characterizes chemical bonding \cite{Szarek2007}. Then, our definition of bond order is
\begin{eqnarray}
b_\varepsilon = \frac{\varepsilon^S_{\tau {\rm AB}}(\vec{r}_L)}{\varepsilon^S_{\tau {\rm HH}}(\vec{r}_L)}. \label{eq:be}
\end{eqnarray}
One should note we normalize by the value of  a H$_2$ molecule calculated at the same level of theory (including method and basis set).

We use Molecular Regional DFT (MRDFT) package \cite{MRDFTv3} to compute these quantities introduced in this section. 
The electronic structure data for the input of the MRDFT package are computed by GAMESS package \cite{GAMESS} with CASSCF method 
in Refs.~\cite{Kurokawa2010} and \cite{Kurokawa2009} (We refer these papers for the details of the CASSCF calculation).
In these calculations, core electrons of transition metals are replaced with Stuttgart-Dresden-Bonn effective core potentials and valence electrons
are represented with a (311111/22111/411/1) basis set \cite{SDD1,SDD2}. 
For C, N, and H, we use cc-pVDZ basis sets \cite{Dunning89} and augmented functions are added to N.
Some part of the visualization is made using 
PyMOL Molecular Viewer program \cite{PyMOL} and Molekel program \cite{Molekel}

\section{Results and discussion} \label{sec:results}

\subsection{Inverted-sandwich-type dinuclear transition metal complexes} \label{sec:istc}
In this section, we discuss the results for the inverted-sandwich-type dinuclear transition metal complexes.
The structures of the complexes to which we apply the electronic stress tensor analysis are shown in Fig.~\ref{fig:istc_be}.
They are model compounds ($\mu$-$\eta^6$:$\eta^6$-C$_6$H$_6$)[M(AIP)]$_2$  (AIPH=(Z)-1-amino-3-imino-prop-1-ene) with
M=V, Cr, Mn and Fe, which have been studied in Ref.~\cite{Kurokawa2010}.
We refer to them as {\bf MB$_{\rm V}$}, {\bf MB$_{\rm Cr}$}, {\bf MB$_{\rm Mn}$} and {\bf MB$_{\rm Fe}$} respectively.
In Ref.~\cite{Kurokawa2010}, it has been shown that the electronic structure of the synthesized complexes,
($\mu$-$\eta^6$:$\eta^6$-C$_6$H$_5$CH$_3$)[Cr(DDP)]$_2$ (DDPH=2-(4-\{(2,6-diisopropylphenyl)imino\}pent-2-ene) \cite{Tsai2007},
($\mu$-$\eta^6$:$\eta^6$-C$_6$H$_5$CH$_3$)[V(DDP)]$_2$ \cite{Tsai2008} and ($\mu$-$\eta^6$:$\eta^6$-C$_6$H$_6$)[Cr(DDP)]$_2$ \cite{Monillas2007},
are very well modeled by {\bf MB$_{\rm Cr}$} and {\bf MB$_{\rm V}$}.
In particular, the observed very high-spin state of septet for the Cr-complex and quintet for the V-complex can be reproduced by these model compounds. 
Also, it has been predicted that the Mn-complex and Fe-complex  have spin state of nonet and singlet respectively. 
We refer Ref.~\cite{Kurokawa2010} for more details of the computational methods and results. 

Using the results of CASSCF calculation carried out by GAMESS package \cite{GAMESS} in Ref.~\cite{Kurokawa2010},
we compute the electronic stress tensor and derived quantities as explained in Sec.~\ref{sec:calc}.
We first show the result of the Lagrange point search in Fig.\ref{fig:istc_be}.
The parts including benzene and transition metals in the complexes are enlarged in  Fig.~\ref{fig:istc_be2}.
We draw a bond line between two atoms when a Lagrange point is found between them and compute 
our energy density bond order $b_\varepsilon$  (eq.~\eqref{eq:be}), which is shown by the number on the bond. 

Most notable feature of this result can be seen at the bonds between benzene and the metal atom. 
We find a Lagrange point for every pair of the metal atom and the C atom in benzene for {\bf MB$_{\rm V}$}, {\bf MB$_{\rm Cr}$} and {\bf MB$_{\rm Mn}$}.
As regards {\bf MB$_{\rm Fe}$}, however, it is found for only two out of six Fe--C pairs.
In contrast, the ligand parts are almost same for all of the complexes. 
Hence, we focus on chemical bonds between metal atoms and benzene in the following. 

We next show the electronic stress tensor and tension.
In Fig.~\ref{fig:stressV}, the case of {\bf MB$_{\rm V}$} is displayed in the plane including a V atom and two C atoms in the benzene. 
As for the electronic stress tensor, we show the largest eigenvalue and corresponding eigenvector in the left panel. 
The sign of the eigenvalue tells whether electrons at a certain point in space feel tensile force (positive eigenvalue) or compressive force (negative eigenvalue), and the eigenvector tells the direction of the force. The largest eigenvalue is considered to give the directionality of the chemical bond. 
In the region between the C atoms, we see that there is a positive eigenvalue region (shown in red) and eigenvectors form a bundle of flow lines, which connects the C atoms.
This pattern is called ``spindle structure" that characterizes the covalent bond \cite{Tachibana2004}.
In the region between the C and V atoms, we can again see a bundle of flow lines that connects two atoms but the eigenvalue is negative (shown in blue).
We call this pattern ``pseudo-spindle structure". 
Now, we turn to the tension vector field, which is shown in the right panel. 
We see that tension vectors basically go out from nuclei in a spherically symmetric manner. 
These vectors sharply change their direction where the vectors from different nuclei meet.
This sharp change creates some surface-like structures in the tension field, and they seem to separate space into subspaces to which atoms in a molecule belong.
This surface is called ``Lagrange surface" \cite{Tachibana2010}.
In the case of {\bf MB$_{\rm V}$}, Lagrange surface always includes a Lagrange point where the tension vanishes.

The cases of {\bf MB$_{\rm Cr}$} and {\bf MB$_{\rm Mn}$} are shown in Figs.~\ref{fig:stressCr} and \ref{fig:stressMn} respectively.
They have a same pattern as {\bf MB$_{\rm V}$}: the spindle structure between the C atoms, the pseudo-spindle structure between the C and the metal atom, and a Lagrange point between each atom pair. 
As for {\bf MB$_{\rm Fe}$}, shown in Figs.~\ref{fig:stressFe} and \ref{fig:stressFe2}, the stress tensor has the same pattern but, as mentioned earlier, there are some metal-C pairs without a Lagrange point.
One may not be able to tell the difference between Figs. \ref{fig:stressV} to \ref{fig:stressMn} and Figs.~\ref{fig:stressFe} or \ref{fig:stressFe2} by visual inspection. In fact, however, the norm of tension vector goes down $10^{-10}$ or smaller at the Lagrange point whereas we cannot find the norm smaller than $10^{-5}$ between Fe(1) and C(13) or C(15). This is the reason why we judge there is no Lagrange point between these atom pairs. 

From the view point of the electronic stress tensor, we may conclude that the V, Cr, and Mn complexes have similar features but the Fe complex is slightly different. 
We now would like to compare this point with the conventional MO analysis.
 The natural orbitals obtained in the study of Ref.~\cite{Kurokawa2010} are shown in Figs.~\ref{fig:NO_Cr}, \ref{fig:NO_Mn} and  \ref{fig:NO_Fe} for {\bf MB$_{\rm Cr}$}, {\bf MB$_{\rm Mn}$} and {\bf MB$_{\rm Fe}$} respectively. 
 We only show the orbitals that are relevant to the metal-benzene bond. 
 We follow the notation of Ref.~\cite{Kurokawa2010} regarding the orbital label so the numbering begins from 3. 
 The occupation numbers are given in the parentheses.
More detailed occupation numbers are given in Table \ref{tab:occupation} which is reproduced from Ref.~\cite{Kurokawa2010} for convenience. 
From Fig.~\ref{fig:NO_Cr} and Fig.~\ref{fig:NO_Mn}, it is easy to specify bonding and anti-bonding orbitals for {\bf MB$_{\rm Cr}$} and  {\bf MB$_{\rm Mn}$} respectively: $\phi_3$ and $\phi_4$ are bonding orbitals and $\phi_{13}$ and $\phi_{14}$ are anti-bonding orbitals. 
The situation is not so clear for {\bf MB$_{\rm Fe}$}.
$\phi_3$ can be specified as bonding and $\phi_{13}$ and $\phi_{14}$ as anti-bonding but $\phi_7$ can be regarded to have bonding orbital feature in addition to $\phi_4$.
In other words, although $\phi_5$ to $\phi_{12}$ of {\bf MB$_{\rm Mn}$} have been assigned to non-bonding-type orbitals in the formal classification \cite{Kurokawa2010}, $\phi_7$ of {\bf MB$_{\rm Fe}$}, in fact, is not 100\% non-bonding and carries some bonding orbital nature.
Such difference between {\bf MB$_{\rm Fe}$} and other three complexes may be reflected in the difference in the Lagrange point pattern we see above.


Finally, we quantify the relative strength of the metal-benzene bond among these complexes by the MO-based bond order and by the energy density-based bond order $b_\varepsilon$.
We are concerned with whether two ways give a consistent result. 
As is done in Ref.~\cite{Kurokawa2010}, bond order can be defined by the half of the difference between the sum of occupation numbers of bonding orbitals and that of anti-bonding orbitals. 
Then, the bond order of {\bf MB$_{\rm V}$}, {\bf MB$_{\rm Cr}$} and {\bf MB$_{\rm Mn}$} is respectively 1.60, 1.34, and 1.07 (see Table \ref{tab:occupation}) and the ratio is $1 : 0.84 : 0.67$.
As for $b_\varepsilon$, it would be reasonable to sum $b_\varepsilon$ of metal-C bonds to obtain the relative strength among {\bf MB$_{\rm V}$}, {\bf MB$_{\rm Cr}$} and {\bf MB$_{\rm Mn}$}. 
$b_\varepsilon$ for the metal-benzene bonds are shown in Fig.~\ref{fig:istc_be2}.
By symmetry, it is sufficient to add three of them. 
Then, the ratio of  {\bf MB$_{\rm V}$} to {\bf MB$_{\rm Cr}$} to {\bf MB$_{\rm Mn}$} regarding the metal-benzene bond is given by  $(0.26+0.24+0.23) : (0.21+0.20+0.20) : (0.20+0.18+0.16) = 0.73 : 0.61 : 0.54$ or $1 : 0.84 : 0.74$. 
This is in reasonable agreement with MO-based bond order. 
The MO bond order for {\bf MB$_{\rm Fe}$} is not calculated because of the ambiguity in the bonding orbital as described earlier, but since the occupation number of anti-bonding orbital $\phi_{14}$ is considerably larger for {\bf MB$_{\rm Fe}$} than for {\bf MB$_{\rm Mn}$}, the metal-benzene bond in {\bf MB$_{\rm Fe}$} should be weaker than that of {\bf MB$_{\rm Mn}$}. 
It is also not capable to compare quantitatively {\bf MB$_{\rm Fe}$} and others using $b_\varepsilon$ due to the different Lagrange point patterns between them. However, the absence of the Lagrange point in {\bf MB$_{\rm Fe}$} suggests that the metal-benzene bond in {\bf MB$_{\rm Fe}$} is weaker than that in {\bf MB$_{\rm Mn}$}.

\subsection{Open-lantern-type transition metal complexes} \label{sec:oltc}
In this section, we discuss the results for the open-lantern-type dinuclear transition metal complexes.
The structures of the complexes to which we apply the electronic stress tensor analysis are shown in Fig.~\ref{fig:oltc_be}.
They are model compounds [M(R$^1$NC(R$^2$)NR$^3$)$_2$]$_2$ 
$({\rm R}^1={\rm R}^2 = {\rm R}^3 = {\rm H})$ with 
M = Cr and Mo, which have been studied in Ref.~\cite{Kurokawa2009}.
We refer to them as {\bf M$_{\rm Cr}$} and {\bf M$_{\rm Mo}$} respectively.\footnote{They are called {\bf M1} and {\bf Mo1} in Ref.~\cite{Kurokawa2009}.}
In Ref.~\cite{Kurokawa2009}, it has been shown that the electronic structure of the synthesized complexes,
 [Cr(R$^1$NC(R$^2$)NR$^3$)$_2$]$_2$ $({\rm R}^1={\rm Et}, {\rm R}^2 = {\rm Me}, {\rm R}^3 = {\rm ^tBu})$,
is very well modeled by {\bf M$_{\rm Cr}$}.
The Cr--Cr distance is calculated to be 1.855\,\AA, which is moderately shorter than the experimental value of 1.960\,\AA.
Also, the bond order of the Cr--Cr bond is calculated to be 2.532, much smaller than the formal bond order of 4.
This can explain the fact that the complex is easy to dissociate into two mononuclear complexes in solution. 
We refer Ref.~\cite{Kurokawa2009} for more details of the computational methods and results. 

Using the results of CASSCF calculation carried out by GAMESS package \cite{GAMESS} in Ref.~\cite{Kurokawa2009},
we compute the electronic stress tensor and derived quantities as explained in Sec.~\ref{sec:calc}.
The result of the Lagrange point search is shown in Fig.\ref{fig:oltc_be}.

We would like to focus on the metal--metal bond for these complexes. 
Let us begin by studying the bond orders. 
The energy density bond order $b_\varepsilon$ for Cr--Cr is 1.42 and that for Mo--Mo is 1.03.
This is contrary to the MO bond order, which is calculated to be 2.532 for Cr--Cr and 3.412 for Mo--Mo \cite{Kurokawa2009}. 
Such a discrepancy is suspected to come from our definition of bond order.
As is defined by Eq.~\eqref{eq:be}, our bond order $b_\varepsilon$ is defined from the energy density evaluated at the Lagrange point. 
It is not difficult to imagine a type of chemical bond that cannot be well characterized by a single point between two atoms. 
This is quite likely to be true for chemical bonds where spatially extended $d$-orbitals are involved; especially, in cases $d_\pi$-$d_\pi$ and $d_\delta$-$d_\delta$ interactions are prominent.

A possible solution to this problem is defining the bond order by integration of the energy density over some surface.
The most natural choice of this surface would be a ``Lagrange surface" \cite{Tachibana2010} that is constructed from a family of lines, which going out from a Lagrange point (if a Lagrange surface includes a Lagrange point).
Namely, we define bond order of the bond between atoms A and B as
\begin{eqnarray}
b_{\varepsilon(S)} =
\frac{ \int_{{\cal S}_{\rm AB}} d^2\sigma \varepsilon^S_\tau(\vec{\sigma})}{ \int_{{\cal S}_{\rm HH}} d^2\sigma \varepsilon^S_\tau(\vec{\sigma})},  \label{eq:beS}
\end{eqnarray}
where ${\cal S}_{\rm AB}$ denotes the Lagrange surface between atoms A and B.
As is the cases of $b_\varepsilon$ (Eq.~\eqref{eq:be}), we normalize by the value of the hydrogen molecule. 

Unfortunately, however, this Lagrange surface is not so easy to define numerically. 
Hence, we instead take the surface integral over the plane that includes a Lagrange point and is perpendicular to the axis connecting two atoms.
Note that such a plane coincides with a Lagrange surface in the case of homonuclear diatomic molecules. 
Also note that in our case of {\bf M$_{\rm Cr}$} and {\bf M$_{\rm Mo}$}, the Lagrange surface between the metal atoms should be very close to such a plane due to the symmetry. 

Another thing we have to determine is the region on the plane over which we integrate the energy density. 
This is because if we integrate all over the plane, we may count energy density which is associated with other bonds. 
We avoid this possibility by integrating only in the region where the eigenvector of the largest eigenvalue of the electronic stress tensor is perpendicular to the plane. 
There could be more than two disconnected regions with such a property, but, of course, we only integrate over the region including the Lagrange point. 
This criterion for the integration region is motivated by the fact that the flow of eigenvectors connecting two atoms is considered to embody a chemical bond. 

Let us now study the concrete cases of {\bf M$_{\rm Cr}$} and {\bf M$_{\rm Mo}$}. 
Fig.~\ref{fig:cs_compare} shows the cross-sections of the Cr--Cr bond and the Mo--Mo bond by the plane discussed earlier.
In detail, energy density distribution and the eigenvector of the largest eigenvalue of the electronic stress tensor are plotted on the plane that includes the Lagrange point of the metal--metal bond and is perpendicular to the bond axis. 
The Lagrange point is located at the origin, and the energy density is normalized by the value at that point.
The energy density is shown by yellow color map and also by the contours.
Since the projection of eigenvectors on this plane is shown by black rods, if the eigenvectors are perpendicular to the plane, they are expressed by dots. 
Then, the regions surrounded by the blue dashed lines (where we cannot see rods) correspond to the regions where eigenvectors are virtually perpendicular to this plane. 
The blue dashed lines are contours on which the perpendicular component of the eigenvector is 0.9. 
As mentioned previously, to calculate the bond order, we shall integrate the energy density over the region surrounded by the blue dashed line, which contains the Lagrange point. 
This is the region located at the central part of the figure. 
Note that this region surrounds the contour for 0.1 of the normalized energy density (thicker red solid line). Therefore, if we integrate the energy density over the region, most of the energy density associated with this bond can be taken into account. 

Here, we report the results of the integration. 
$b_{\varepsilon(S)}$ for Cr--Cr is 2.92 and that for Mo--Mo is 3.13.
Before integration, namely in terms of $b_\varepsilon$, Cr--Cr is calculated to be stronger than Mo--Mo but after integration, in terms of $b_{\varepsilon(S)}$, Cr--Cr is calculated to be weaker than Mo--Mo, which is consistent with the MO-based bond order calculation. 
The relative strength measured by $b_\varepsilon$ turns out to be reversed from the one measured by $b_{\varepsilon(S)}$ because the energy density distribution in Mo--Mo is broader than that of Cr--Cr (see Fig.~\ref{fig:cs_compare}).
This is again consistent with the MO analysis of Ref.~\cite{Kurokawa2009}, which has concluded the spatial extension of $d$-orbitals of Cr is less than that of Mo and with earlier literature \cite{Frenking2000}. 

In addition, we check that relative strength of the metal-benzene bond among  {\bf MB$_{\rm V}$}, {\bf MB$_{\rm Cr}$} and {\bf MB$_{\rm Mn}$} which have been analyzed in Sec.~\ref{sec:istc} does not change if we use $b_{\varepsilon(S)}$ instead of $b_{\varepsilon}$. We have seen in Sec.~\ref{sec:istc} that  the ratio of  {\bf MB$_{\rm V}$} to {\bf MB$_{\rm Cr}$} to {\bf MB$_{\rm Mn}$} regarding the metal-benzene bond is $1 : 0.84 : 0.74$ when we use $b_{\varepsilon}$. When we use  $b_{\varepsilon(S)}$, the ratio is $1:0.89:0.76$, preserving the same ordering. This is consistent with the fact that these bonds do not involve spatially extended $d$-orbitals.

We would like to end this section by examining the electronic stress tensor of the metal-metal bond. 
The results are shown in Figs.~\ref{fig:stressCrCr} and \ref{fig:stressMoMo}.
They are drawn on the plane including two metal atoms and the angle between the plane of Fig.~\ref{fig:cs_compare} is 90$^\circ$.
As for the Cr--Cr bond, we see flow lines that connect the Cr atoms with positive eigenvalue region, that is, a spindle structure.
As for the Mo--Mo bond, the Mo atoms are similarly connected by the flow lines of eigenvectors and we see the positive eigenvalue region but it is not simply connected. 
In particular, it takes negative value at the Lagrange point. 
We may say this is a spindle structure but it partly has some feature of a pseudo-spindle structure. 

To discuss the negative eigenvalue region found in the Mo--Mo bond, it is instructive to look at the C--C bonds of C$_2$H$_6$, C$_2$H$_4$ and C$_2$H$_2$. This has been investigated in Ref.~\cite{Tachibana2005} but we show the results in Fig.~\ref{fig:stressC2} for convenience. As we can see there, while C$_2$H$_6$ and C$_2$H$_4$ have spindle structures, C$_2$H$_2$ has a pseudo-spindle structure.\footnote{
If calculated at the cc-pVDZ level, we see small regions with positive eigenvalue just like those of the Mo--Mo bond in Fig~\ref{fig:stressMoMo}. Since such regions do not appear at 6-31G(d,p), cc-pVTZ and cc-pVQZ levels, we believe they are numerical artifacts in the case of C$_2$H$_2$.
}
The negative eigenvalue of C$_2$H$_2$ is caused by the compressive stress nearby the C nuclei. 
In general, the stress tensor has a large negative eigenvalue in radial direction in neighborhood of a nucleus due the dominance of the attractive Coulomb force. 
In the case of C$_2$H$_2$, the bond length is too short that the internuclear region is immersed under the atomic compressive stress \cite{Tachibana2005}. 
If we regard C$_2$H$_6$, C$_2$H$_4$ and C$_2$H$_2$ as a series which changing from a spindle structure to a pseudo-spindle structure, the Mo--Mo bond in {\bf M$_{\rm Mo}$} may correspond to the stage between C$_2$H$_4$ and C$_2$H$_2$.
Of course, since they have totally different shell structures, the direct comparison does not make sense. 
However, it stimulates us to look for other compounds with stronger/weaker Mo--Mo bonds and see whether they produce pseudo-spindle/spindle structures. 
This is similar for the Cr--Cr bond.
Whether stronger Cr--Cr bond than that of M$_{\rm Cr}$ produces (partly) pseudo-spindle structure is a very interesting question to ask. 
Now, final comments are in order. 
We have just mentioned that when considering the series which changes from a spindle structure to a pseudo-spindle structure, it is non-sense to discuss different types of atoms on equal footing because of difference in shell structures.  
Meanwhile, we have shown that the energy density, which is dynamically well-defined quantity, is capable of showing the Mo--Mo bond is stronger than the Cr--Cr bond. 
Then, the spindle/pseudo-spindle structure series may be discussed in a unified manner using the energy density and stress tensor. 
To do this, it is also essential to clarify how the stress tensor changes as the shells pile up and its effect on chemical bonds.

\section{Summary} \label{sec:summary}
In this paper, we have investigated the electronic structure of two types of transition metal complexes, the inverted-sandwich-type and open-lantern-type, by the electronic stress tensor. 
In particular, the bond order $b_\varepsilon$ measured by the energy density which is defined from the electronic stress tensor has been studied and compared with the conventional MO-based bond order. 
We have also studied the patterns found in the largest eigenvalue of the stress tensor and corresponding eigenvector field, the ``spindle structure" and ``pseudo-spindle structure". They are both defined by a bundle of flow lines formed by the eigenvectors which connects two atoms and the former has the positive eigenvalue while the latter has the negative eigenvalue. 

As for the inverted-sandwich-type complex, we have investigated V, Cr, Mn, and Fe complexes. 
Our bond order $b_\varepsilon$ calculation has shown that relative strength of the metal-benzene bond among V, Cr, and Mn complexes is ${\rm V > Cr > Mn}$ which turned out to be same as the MO-based bond order as was found in Ref.~\cite{Kurokawa2010}. 
The Fe complex has not been investigated in this context due to the different pattern of the Lagrange points (on which the energy density is computed to define $b_\varepsilon$). 
This is in a sense also consistent with the MO analysis because the bonding/non-bonding orbital assignment for the Fe complex 
was rather ambiguous and not as clear as that for the other three complexes. 
We have also studied the eigenvector pattern of the largest eigenvalue of the stress tensor. 
The bond between the metal atom and C atom of benzene are characterized by the pseudo-spindle structure for all of the complexes. 
It was found that some of the pseudo-spindle structures were not associated with a Lagrange point. 

Regarding the open-lantern-type complex, we have investigated Cr and Mo complexes.
In this case, $b_\varepsilon$ calculation has shown that relative strength of the metal-metal bond between Cr and Mo complexes is ${\rm Cr > Mo}$ which is reversed order to the MO-based bond order calculated in Ref.~\cite{Kurokawa2009}. 
Suspecting that $b_\varepsilon$, which measure the energy density at a single point, is not appropriate for a bond involving spatially extended $d$-orbitals, we have proposed a modified definition of the energy density-based bond order, $b_{\varepsilon(S)}$, Eq.~\eqref{eq:beS}.
This new definition measures the energy density integrated over the ``Lagrange surface" between two atoms and is able to take into account spatial extent of the energy density. 
Actually, using $b_{\varepsilon(S)}$, the relative strength of Cr--Cr and Mo--Mo was calculated to be ${\rm Cr < Mo}$ which is consistent with the MO-based bond order.
Finally, we have studied the eigenvector pattern in the regions of Cr--Cr and Mo--Mo bonds.
The Cr--Cr was found to be characterized by a positive eigenvalue region, which is spindle structure, while the Mo--Mo has shown both positive and negative regions. It seems to be characterized by a spindle structure but also has a pseudo-spindle nature. 

Although our study here was carried out for limited types of transition metal complexes, we were able to gain new insight into the eigenvector field pattern of the largest eigenvalue of the electronic stress tensor and the Lagrange point patterns, which had not been found in our previous studies of $s$- or $p$-block compounds. 
We also have confirmed that our energy density-based bond order can properly describe the relative strength of chemical bonds involving $d$-orbitals by the surface integration. 
Further study of transition metal compounds in particular including metal-metal bonds will lead us to deepen our understanding of the electronic stress tensor and nature of $\delta$-bonding.

\noindent 
\section*{Acknowledgments}
Theoretical calculations were partly carried out using Research Center for
 Computational Science, Okazaki, Japan.
This research work is supported partly by Collaborative Research Program for Young Scientists of ACCMS and IIMC from Kyoto University. 
This work is supported partly by Grant-in-Aid for Scientific research (No.~22550011),
Grant-in-Aid for Specially Promoted Research (No.~22000009), and Grand Challenge Project (IMS, Okazaki, Japan) 
from the Ministry of Education, Culture, Sports, Science and Technology, Japan.



\newpage

\begin{table}
\caption{Occupation numbers of the natural orbitals which are relevant to the metal-benzene bonds of the inverted-sandwich-type complexes. After Table 2 of Ref.~\cite{Kurokawa2010}.}
\begin{center}
\begin{tabular}{|c|c|c|c|c|}
\hline
 & {\bf MB$_{\rm V}$} & {\bf MB$_{\rm Cr}$} & {\bf MB$_{\rm Mn}$} & {\bf MB$_{\rm Fe}$}  \\
\hline
\hline
$\phi_3$ & 1.7968 & 1.6660 & 1.5312 & 1.8710 \\
$\phi_4$ & 1.8018 & 1.6639 & 1.5391 & 1.8633 \\
$\phi_5$ & 1.0000 & 1.0031 & 1.0001 & 1.0045 \\
$\phi_6$ & 0.9980 & 1.0020 & 1.0000 & 0.9956 \\
$\phi_7$ & 0.9946 & 0.9977 & 1.0010 & 1.6929 \\
$\phi_8$ &              &  0.9967 & 0.9996 & 1.5952 \\
$\phi_9$ &              & 1.0000 & 1.0000 & 1.0445 \\
$\phi_{10}$ &         &  1.0000 & 1.0000 & 0.9571 \\
$\phi_{11}$ &         &               & 1.0000 & 1.0450 \\ 
$\phi_{12}$ &          &             &  1.0000 & 0.9564 \\
$\phi_{13}$ & 0.2052 & 0.3344 & 0.4676 & 0.4336 \\
$\phi_{14}$ & 0.2029 & 0.3362 & 0.4615 & 0.5409 \\
\hline
\end{tabular}
\end{center}
\label{tab:occupation}
\end{table}


\ifFIG

\newpage

\begin{figure}
\begin{center}
\vspace{-2cm}
\includegraphics[width=11cm]{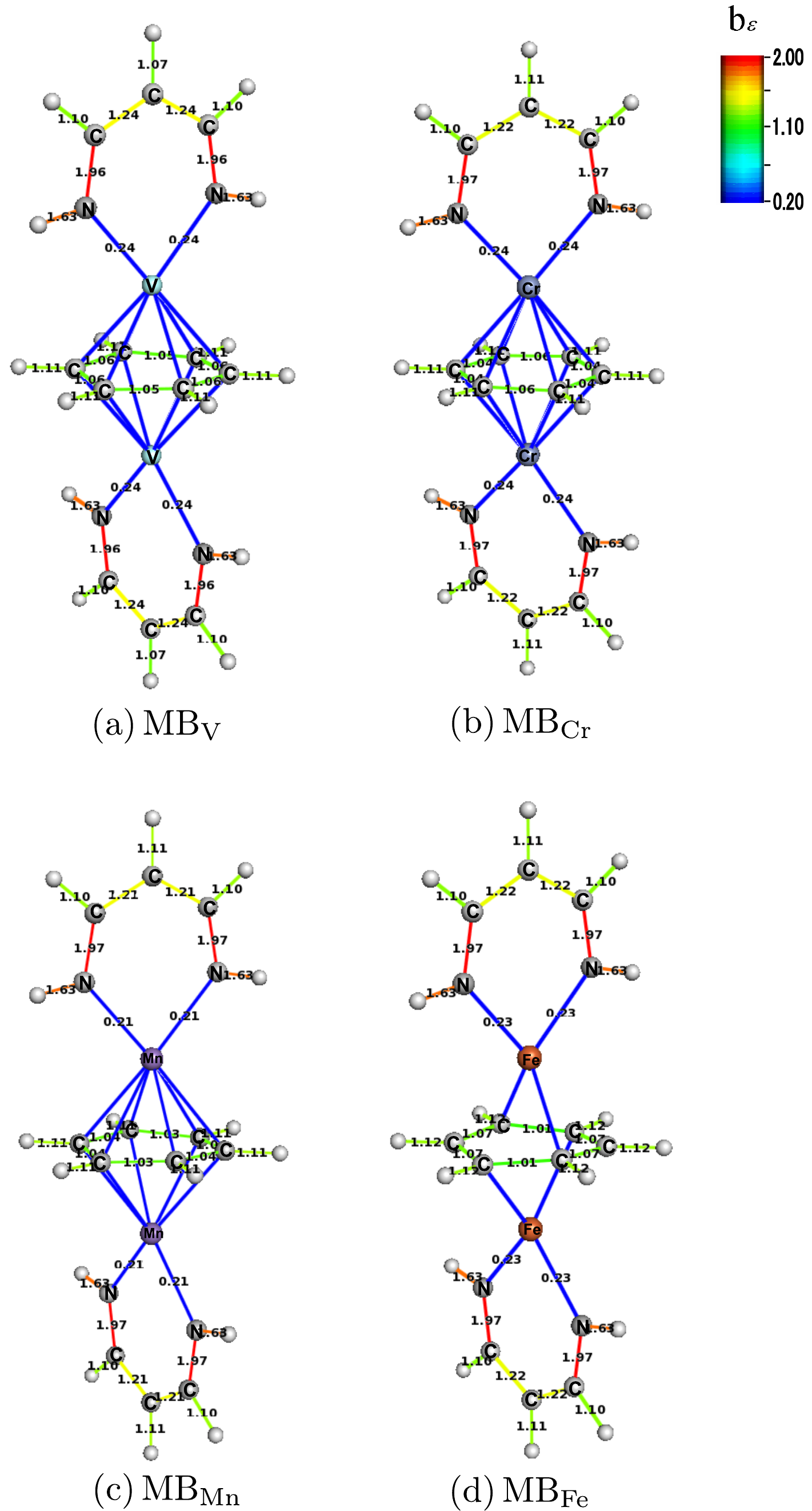}
\caption{Structures and bond order $b_\varepsilon$ for the inverted-sandwich-type dinuclear transition metal complexes: (a) {\bf MB$_{\rm V}$}, (b) {\bf MB$_{\rm Cr}$}, (c) {\bf MB$_{\rm Mn}$} and  (d) {\bf MB$_{\rm Fe}$}.
A bond line is drawn between two atoms when a Lagrange point is found between them and our energy density based bond order $b_\varepsilon$  (eq.~\eqref{eq:be}) is shown by color and the number on the bond.  }
\label{fig:istc_be}
\end{center}
\end{figure}

\begin{figure}
\begin{center}
\includegraphics[width=15cm]{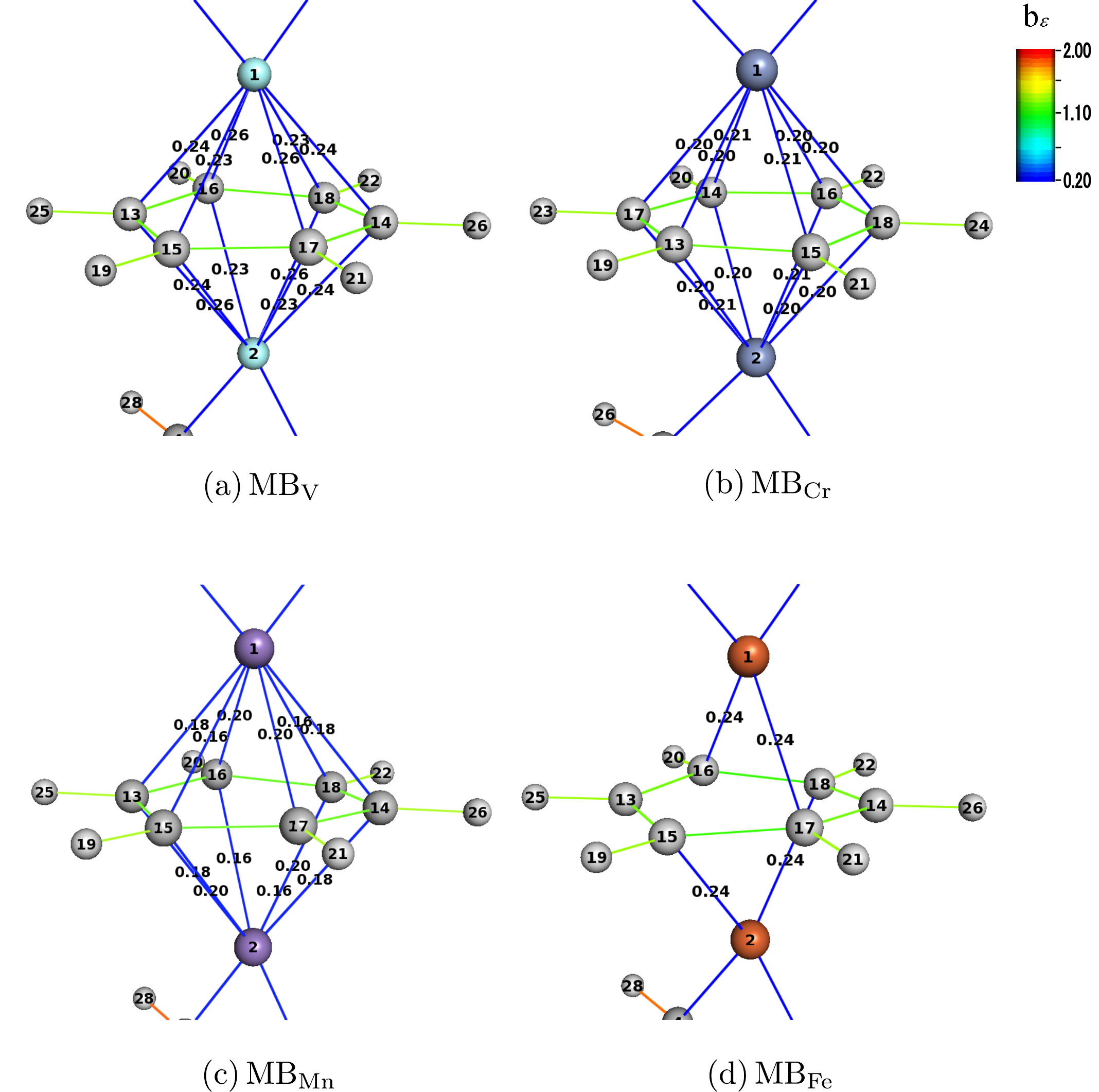}
\caption{Close-ups of Fig.~\ref{fig:istc_be} at the benzene and metal atoms:  (a) {\bf MB$_{\rm V}$}, (b) {\bf MB$_{\rm Cr}$}, (c) {\bf MB$_{\rm Mn}$} and  (d) {\bf MB$_{\rm Fe}$}.}
\label{fig:istc_be2}
\end{center}
\end{figure}

\begin{figure}
\begin{center}
\includegraphics[width=16cm]{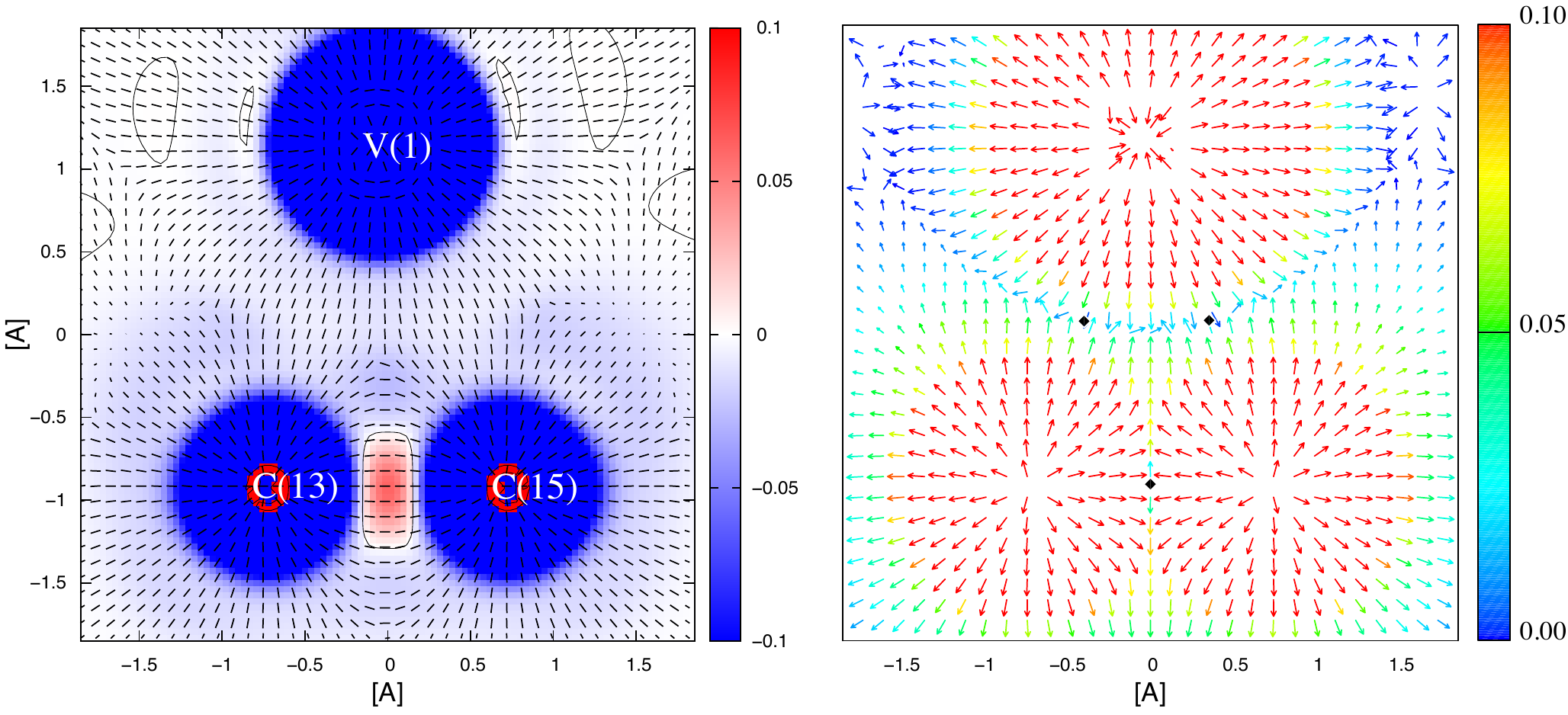}
\caption{The left panel shows the largest eigenvalue of the stress tensor (color map) and corresponding eigenvector (black rods) of {\bf MB$_{\rm V}$}
on the plane including the labelled atoms (see Fig.~\ref{fig:istc_be2} for the number in the label). 
As for the eigenvectors, the projection on this plane is plotted. 
The black solid line denotes a contour where the eigenvalue is zero.
The right panel shows the tension on the same plane. 
The tension vectors are normalized and the projection on this plane is plotted.
The norm is expressed by the color of the arrows.
Also, the locations of the Lagrange point are marked by the black diamonds. 
}		
\label{fig:stressV}
\end{center}
\end{figure}

\begin{figure}
\begin{center}
\includegraphics[width=16cm]{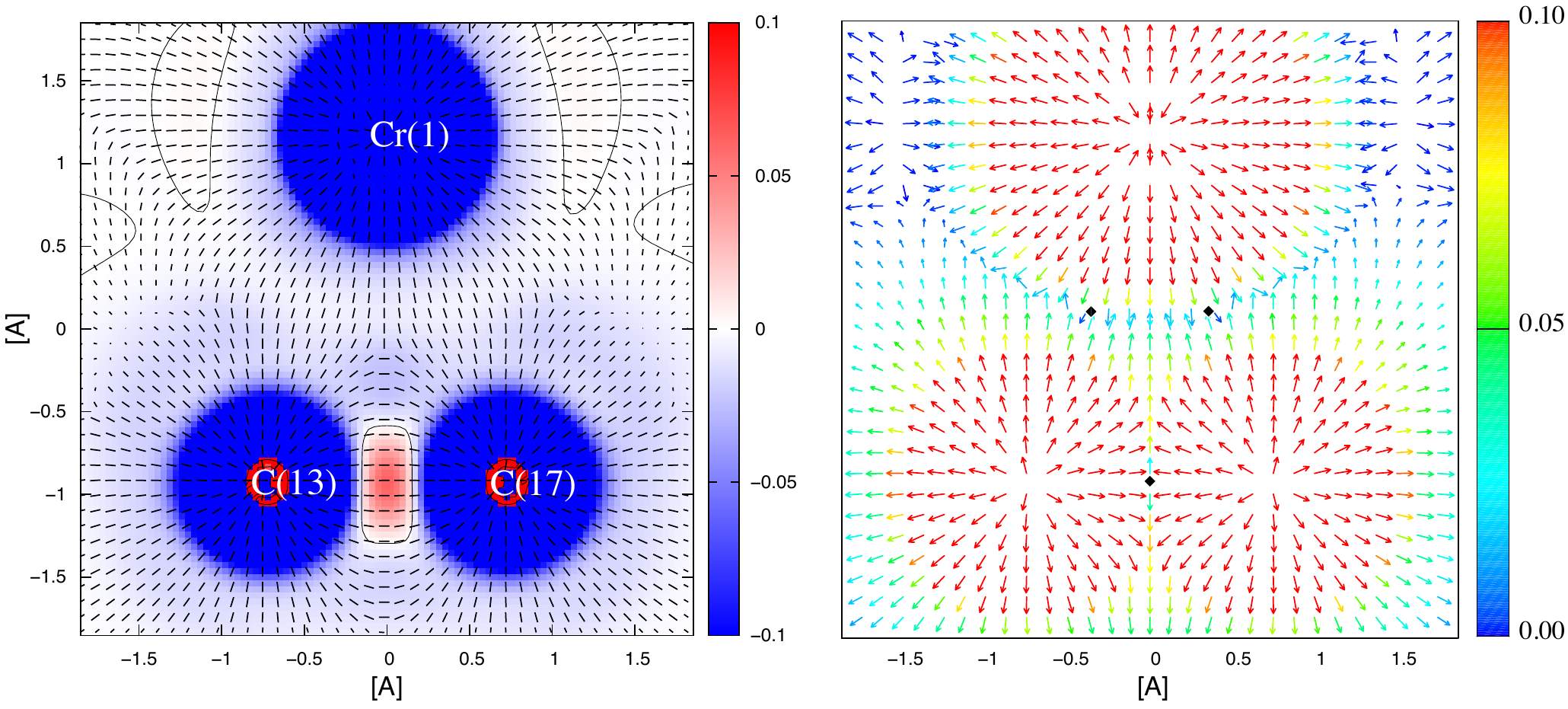}
\caption{The stress tensor and tension of {\bf MB$_{\rm Cr}$} plotted in the same manner as Fig.~\ref{fig:stressV}.}		
\label{fig:stressCr}
\end{center}
\end{figure}

\begin{figure}
\begin{center}
\includegraphics[width=16cm]{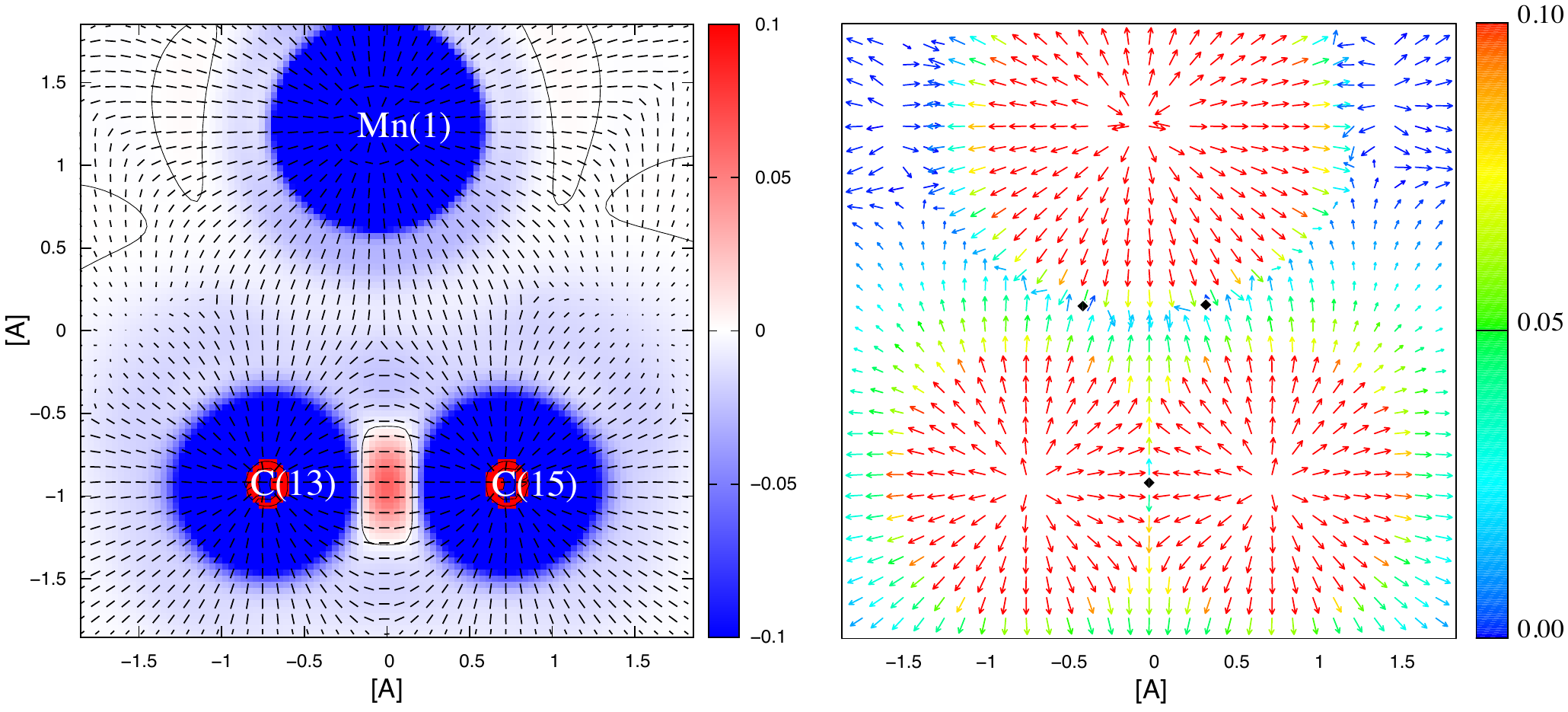}
\caption{The stress tensor and tension of {\bf MB$_{\rm Mn}$} plotted in the same manner as Fig.~\ref{fig:stressV}.}		
\label{fig:stressMn}
\end{center}
\end{figure}

\begin{figure}
\begin{center}
\includegraphics[width=16cm]{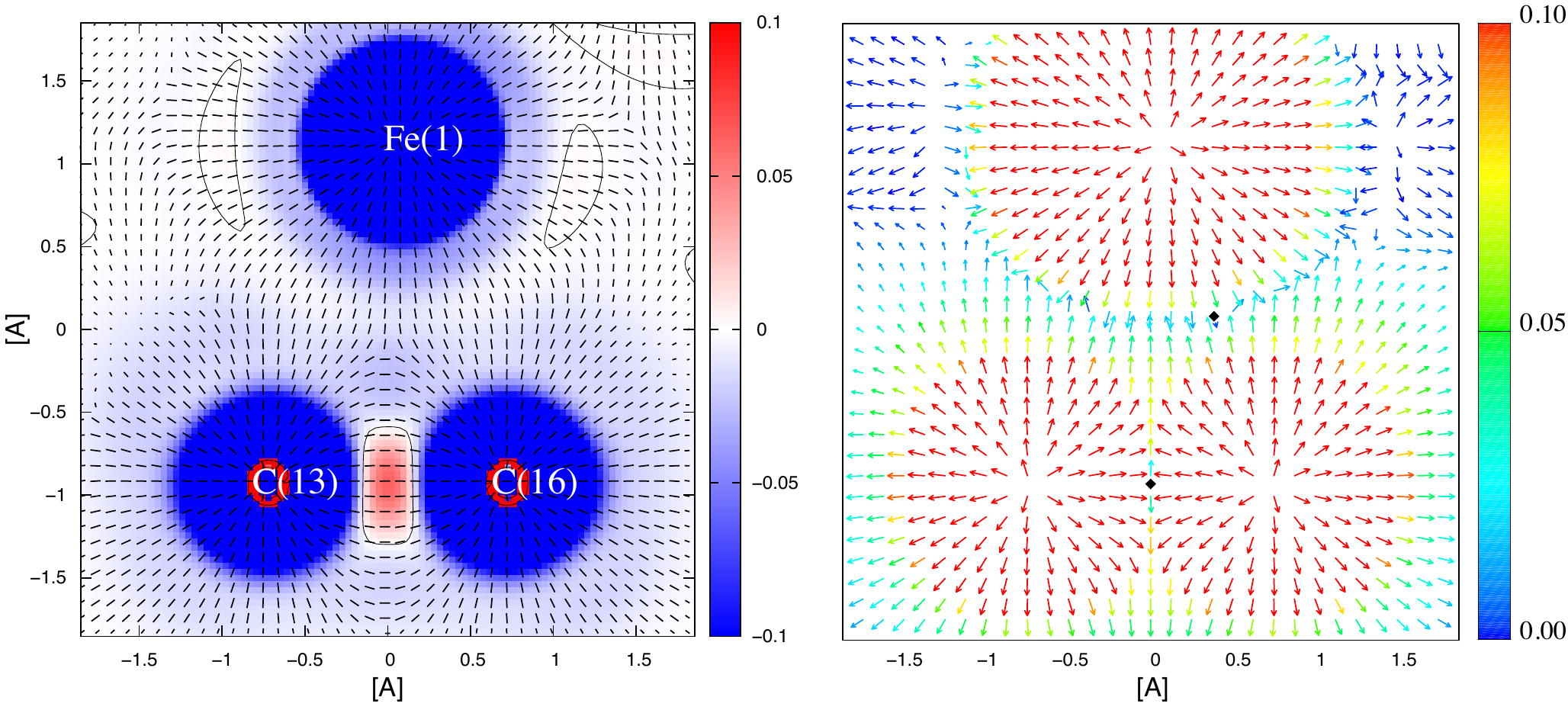}
\caption{The stress tensor and tension of {\bf MB$_{\rm Fe}$} plotted in the same manner as Fig.~\ref{fig:stressV}. 
There is no Lagrange point between Fe(1) and C(13).}		
\label{fig:stressFe}
\end{center}
\end{figure}

\begin{figure}
\begin{center}
\includegraphics[width=16cm]{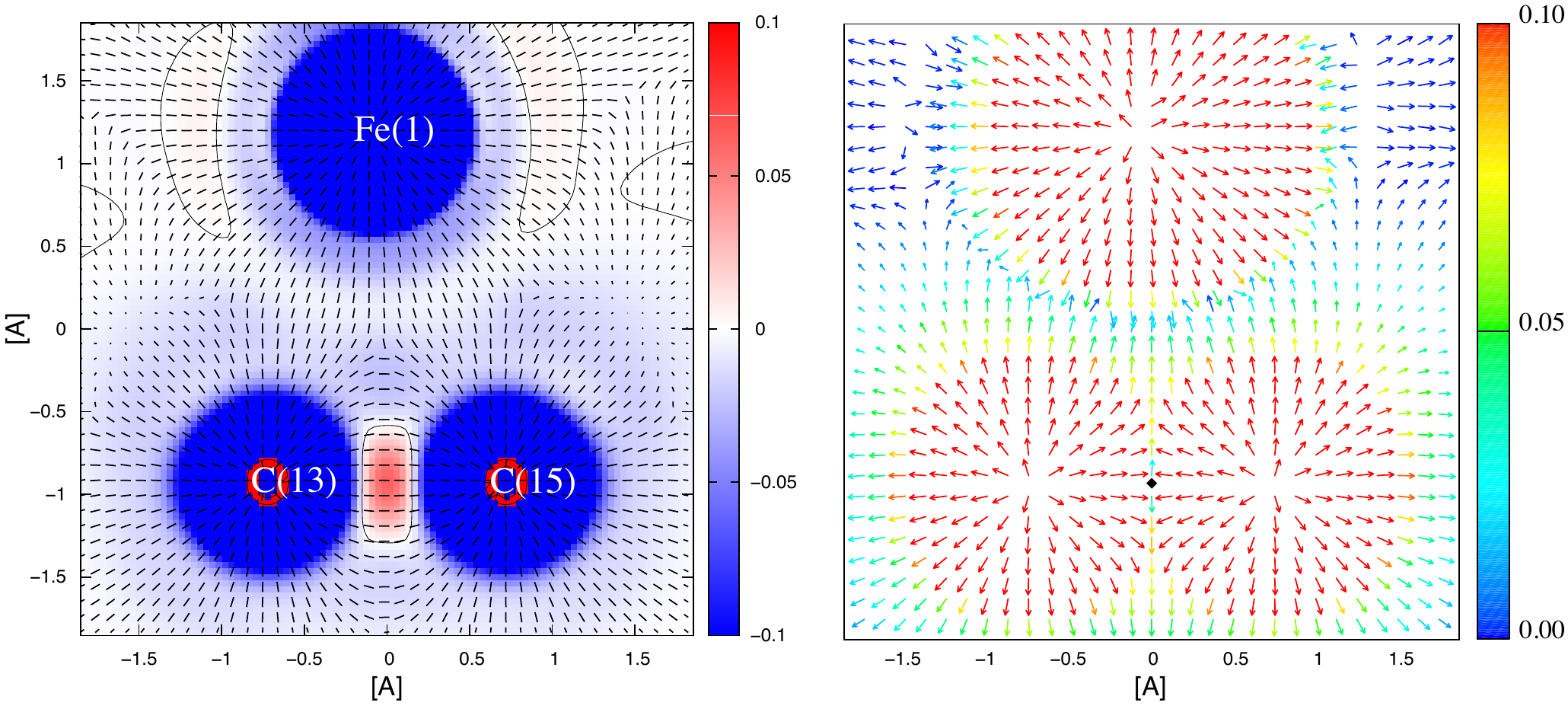}
\caption{The stress tensor and tension of {\bf MB$_{\rm Fe}$} plotted in the same manner as Fig.~\ref{fig:stressV} on the different plane from Fig.~\ref{fig:stressFe}. 
There is no Lagrange point between Fe(1) and C(13)/C(15). }		
\label{fig:stressFe2}
\end{center}
\end{figure}


\begin{figure}
\begin{center}
\includegraphics[width=12cm]{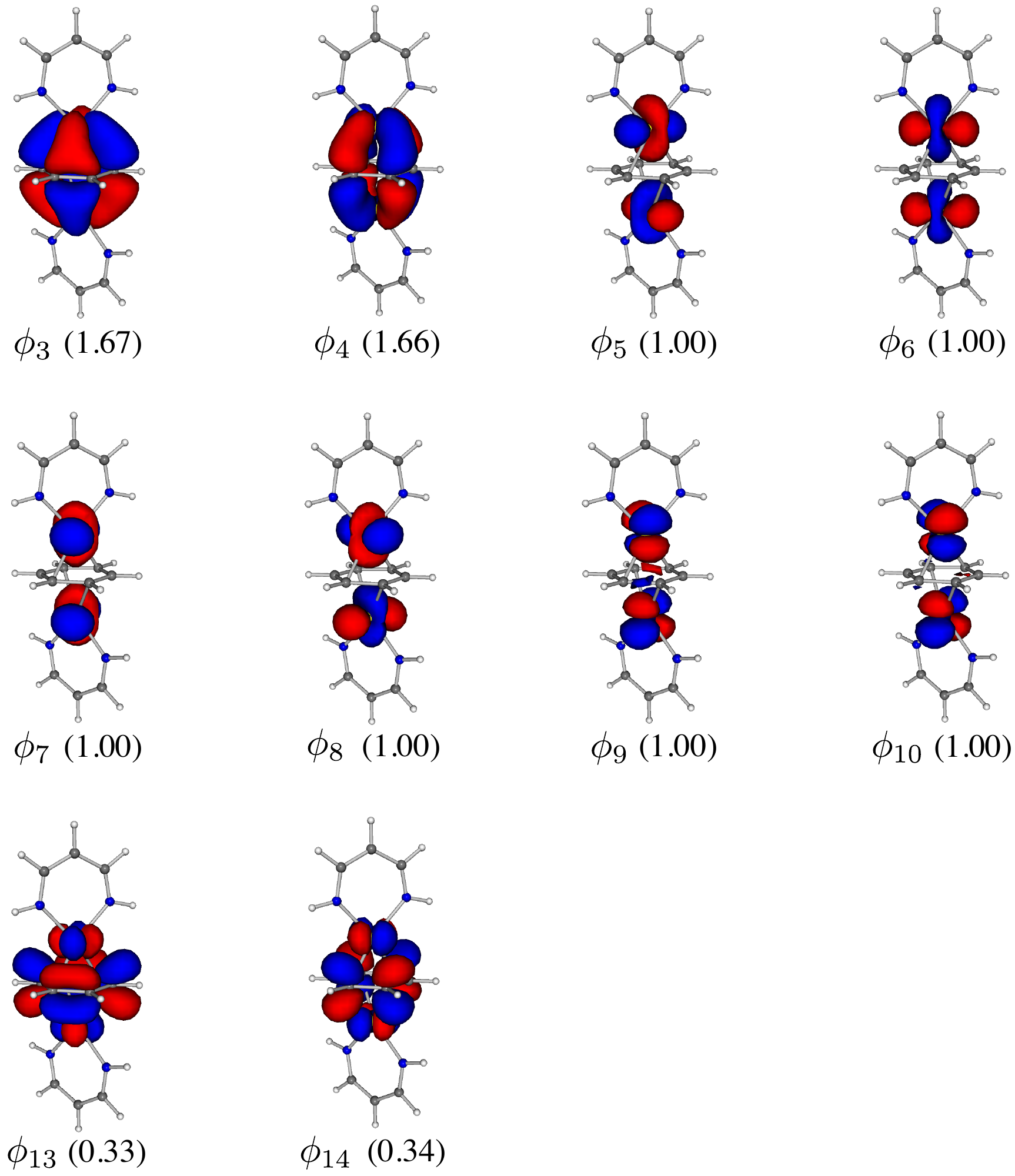}
\caption{Natural orbitals which are relevant to the metal-benzene bonds of {\bf MB$_{\rm Cr}$}. Their occupation numbers are shown in parentheses. }
\label{fig:NO_Cr}
\end{center}
\end{figure}

\begin{figure}
\begin{center}
\includegraphics[width=12cm]{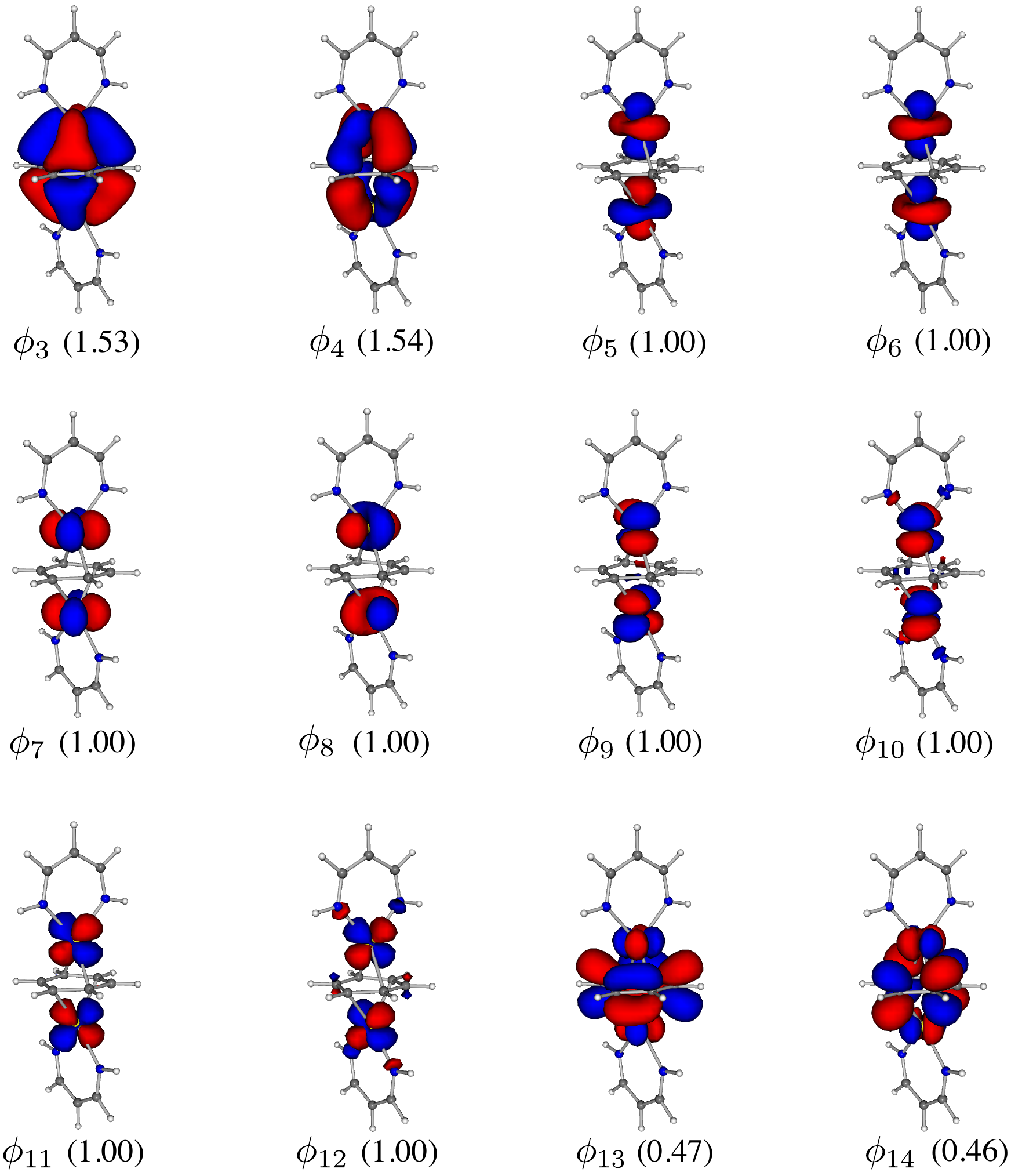}
\caption{Natural orbitals which are relevant to the metal-benzene bonds of {\bf MB$_{\rm Mn}$}. 
Their occupation numbers are shown in parentheses. }
\label{fig:NO_Mn}
\end{center}
\end{figure}

\begin{figure}
\begin{center}
\includegraphics[width=12cm]{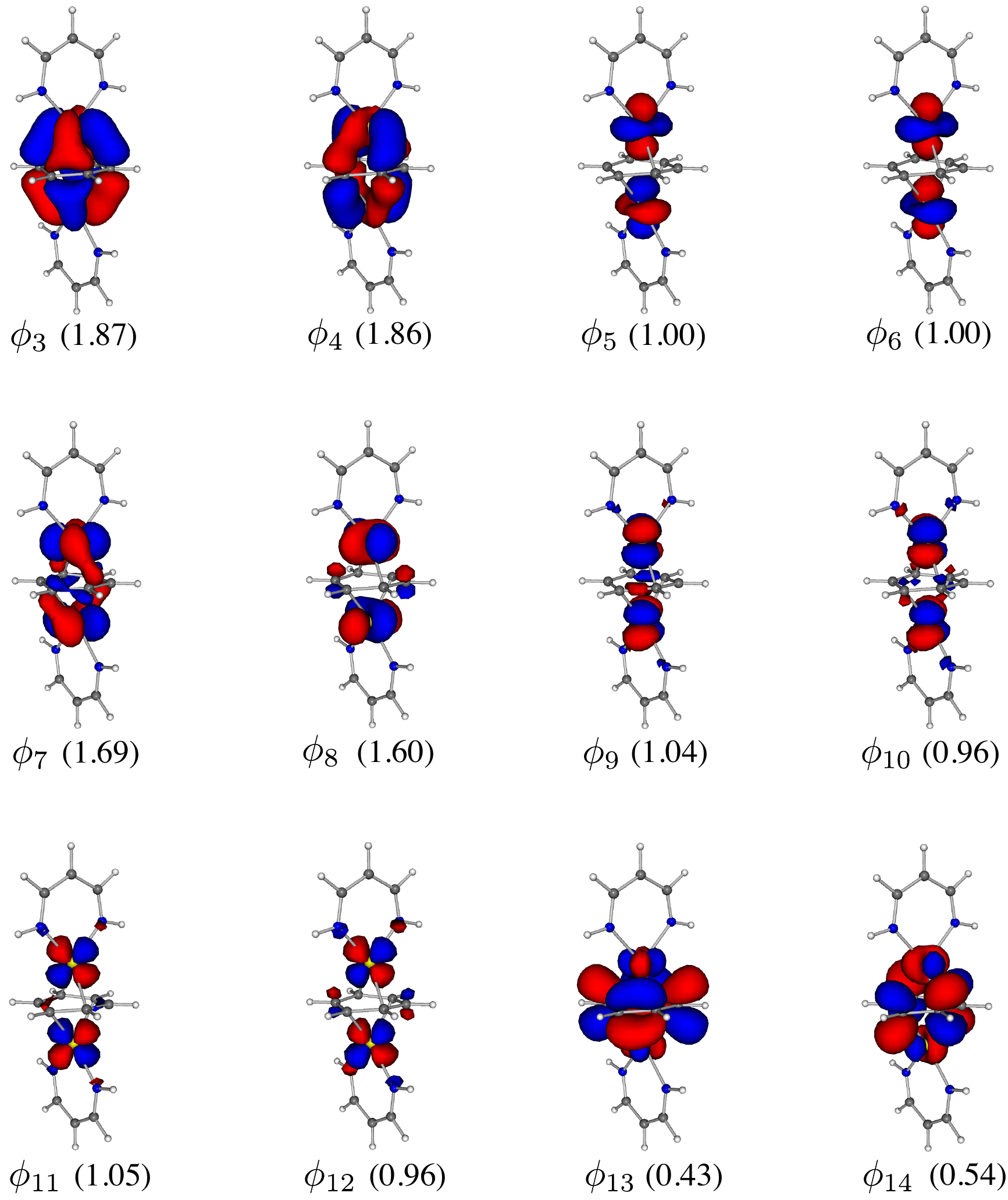}
\caption{Natural orbitals which are relevant to the metal-benzene bonds of {\bf MB$_{\rm Fe}$}. Their occupation numbers are shown in parentheses. }
\label{fig:NO_Fe}
\end{center}
\end{figure}

\begin{figure}
\begin{center}
\includegraphics[width=14cm]{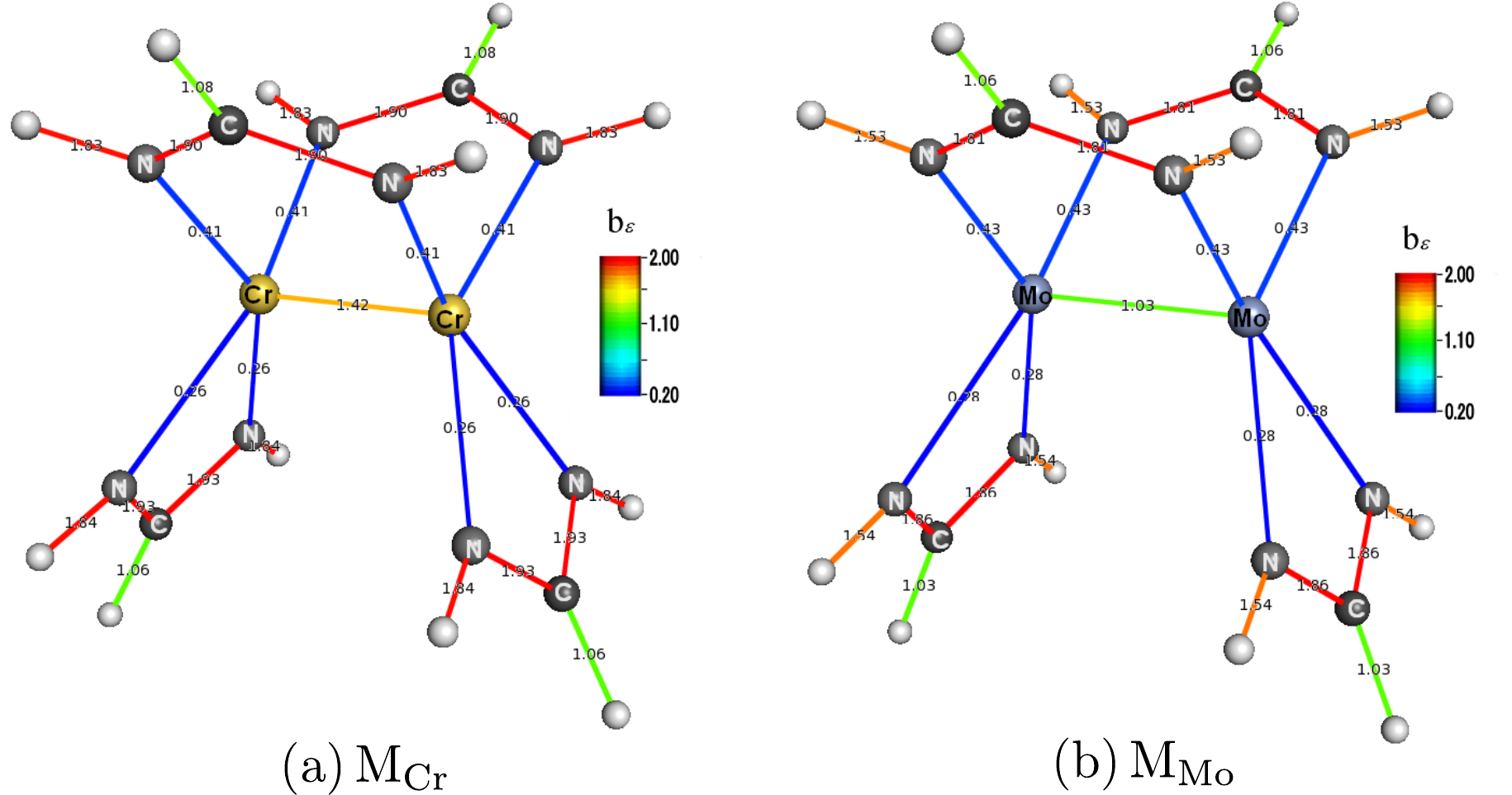}
\caption{Structures and bond order $b_\varepsilon$ for the open-lantern-type dinuclear transition metal complexes: (a) {\bf M$_{\rm Cr}$} and (b) {\bf M$_{\rm Mo}$}.
A bond line is drawn between two atoms when a Lagrange point is found between them and our energy density based bond order $b_\varepsilon$  (eq.~\eqref{eq:be}) is shown by color and the number on the bond. }
\label{fig:oltc_be}
\end{center}
\end{figure}

\begin{figure}
\begin{center}
\includegraphics[width=15cm]{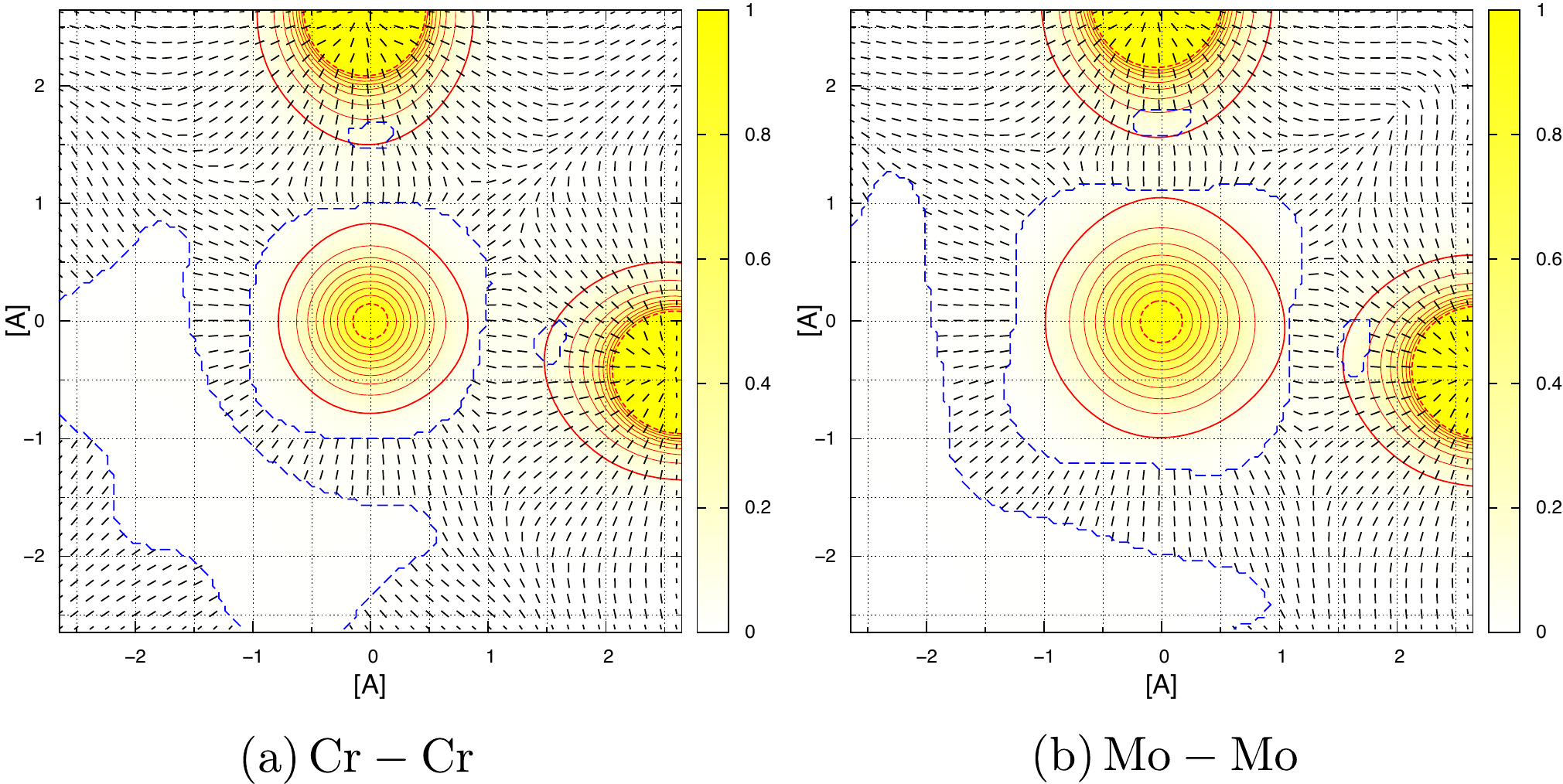}
\caption{
Energy density distribution and the eigenvector of the largest eigenvalue of the electronic stress tensor of (a) {\bf M$_{\rm Cr}$} and (b) {\bf M$_{\rm Mo}$}. 
They are plotted on the plane which includes the Lagrange point of the metal--metal bond and is perpendicular to the bond axis. 
The Lagrange point is located at the origin.
The yellow color map shows the energy density which is normalized by the value at the Lagrange point.
The energy density is also shown by the contours.
The thicker red solid line is for 0.1 of the normalized energy density and the thicker red dashed line is for 0.9.
The other contours denote values at intervals of 0.1 between them. 
As for the eigenvectors, the projection on this plane is shown by black rods. 
Then, the regions without rods surrounded by the blue dashed lines correspond to the regions where eigenvectors are virtually perpendicular to this plane. 
The blue dashed lines are contours on which the perpendicular component of the eigenvector is 0.9. 
 }
\label{fig:cs_compare}
\end{center}
\end{figure}

\begin{figure}
\begin{center}
\includegraphics[width=16cm]{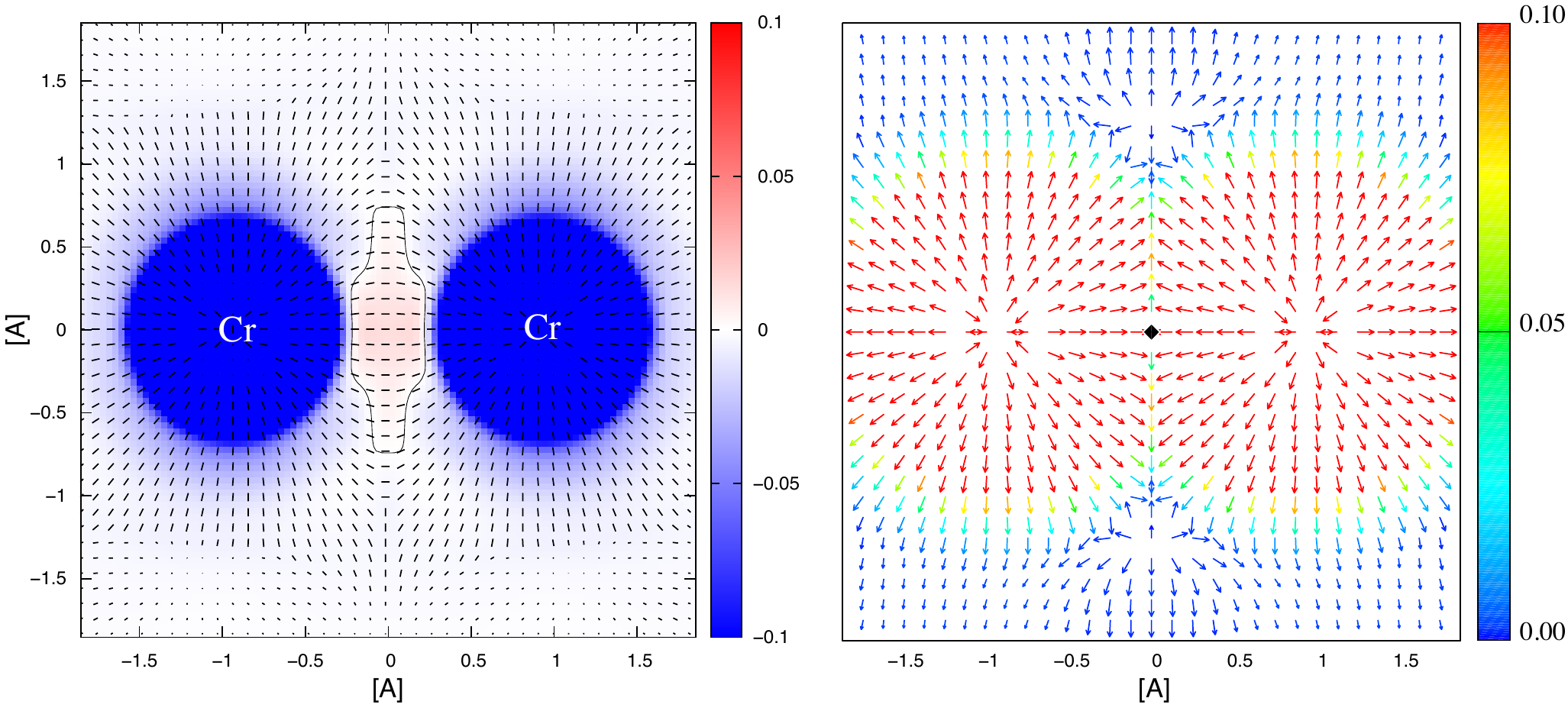}
\caption{The stress tensor and tension of {\bf M$_{\rm Cr}$} plotted in the same manner as Fig.~\ref{fig:stressV}.}		
\label{fig:stressCrCr}
\end{center}
\end{figure}

\begin{figure}
\begin{center}
\includegraphics[width=16cm]{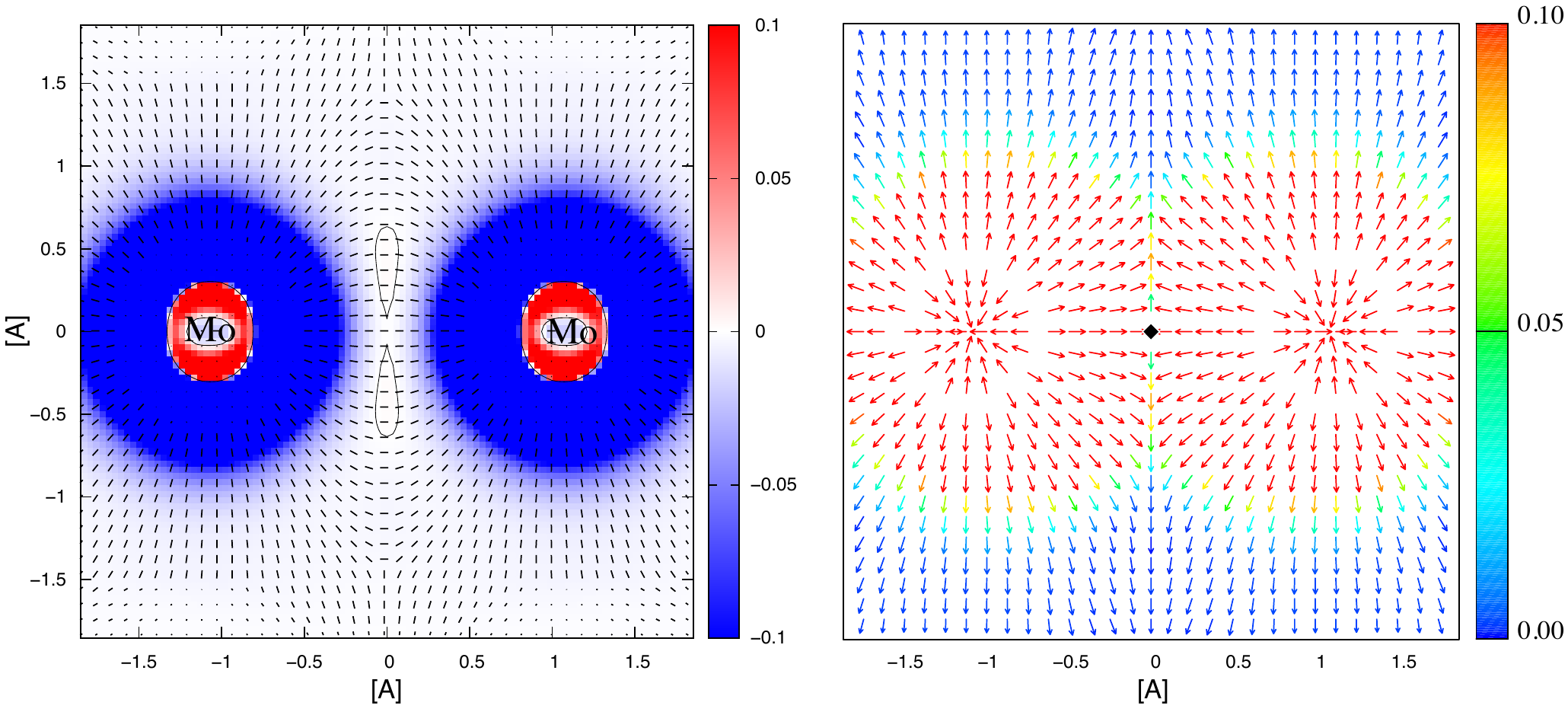}
\caption{The stress tensor and tension of {\bf M$_{\rm Mo}$} plotted in the same manner as Fig.~\ref{fig:stressV}.}		
\label{fig:stressMoMo}
\end{center}
\end{figure}

\begin{figure}
\begin{center}
\includegraphics[width=7cm]{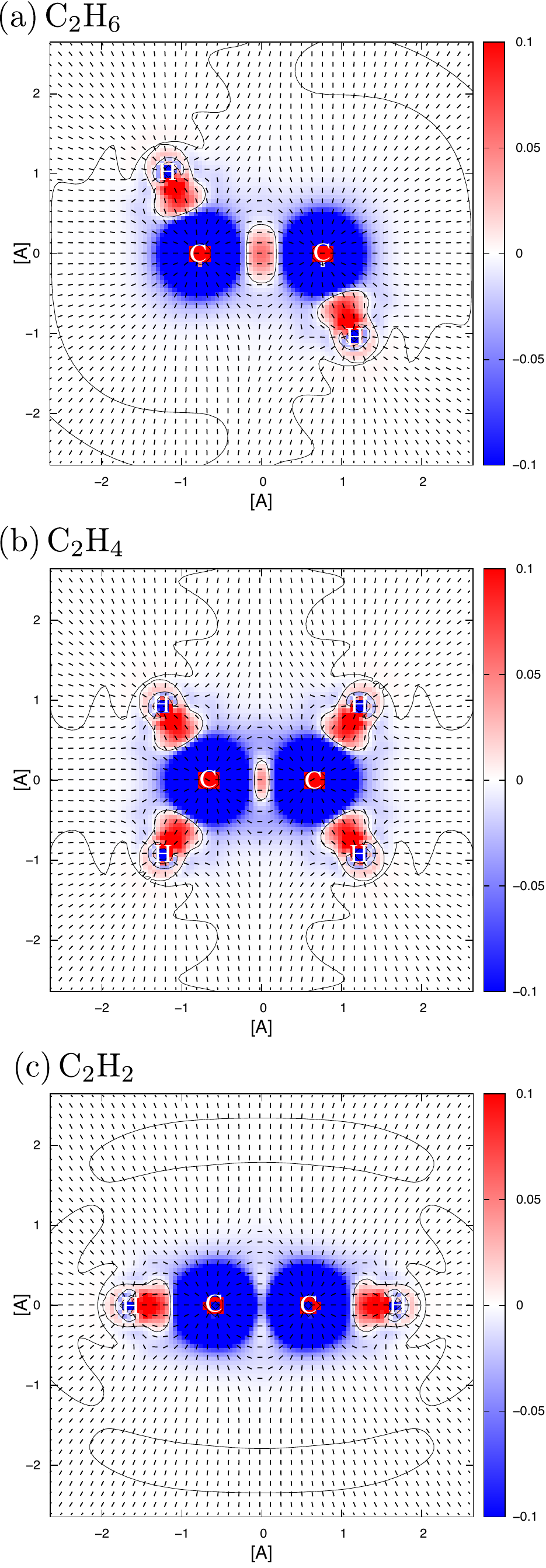}
\caption{The stress tensor of (a) C$_2$H$_6$,  (b) C$_2$H$_4$ and (c) C$_2$H$_2$ plotted in the same manner as Fig.~\ref{fig:stressV}. Calculated by HF/cc-pVQZ \cite{Dunning89}. }		
\label{fig:stressC2}
\end{center}
\end{figure}

\fi

\end{document}